\documentstyle[12pt,epsfig,a4]{article} 
\newcommand{\vs}{\vspace{-0.25cm}}
\begin{document} 
\begin{center}
\large{\bf Twice-iterated boson-exchange scattering amplitudes}

N. Kaiser\\

\smallskip

{\small Physik Department T39, Technische Universit\"{a}t M\"{u}nchen,
    D-85747 Garching, Germany}

\end{center}

\begin{abstract} We calculate at two-loop order the complex-valued scattering 
amplitude related to the twice-iterated scalar-isovector boson-exchange between
nucleons. In comparison to the once-iterated boson-exchange amplitude it shows 
less dependence on the scattering angle. We calculate also the iteration of
the (static) irreducible one-loop potential with the one-boson exchange and 
find similar features. Together with the irreducible three-boson exchange 
potentials and the two-boson exchange potentials with vertex corrections,
which are also evaluated analytically, our results comprise all
nonrelativistic contributions from scalar-isovector boson-exchange at one- and 
two-loop order. The applied methods can be straightforwardly adopted to
the pseudoscalar pion with its spin- and momentum-dependent couplings to the
nucleon.         
\end{abstract}

\medskip
PACS: 12.20.Ds, 12.38.Bx, 13.75.Cs, 21.30.Cb.

\bigskip


\bigskip

Over the last years, effective field theory (in particular chiral perturbation 
theory) has been successfully applied to the two-nucleon system at low and 
intermediate energies \cite{evgeni1,evgeni2,entem}. The strategy is to 
construct the long- and medium-range NN-potential systematically from one-,
two- and three-pion exchanges and to represent the short-distance dynamics by
adjustable contact interactions. In order to obey unitarity of the S-matrix
the NN-potential must then be iterated to all orders, e.g., by solving
the Lippmann-Schwinger integral equation. In practice this unitarization
procedure introduces additional cut-offs or off-shell form factors in order to
eliminate high-momentum components from the chiral NN-potential. However,
there exist also purely algebraic methods for unitarization, such as the
inverse amplitude expansion or the N/D-method, which do not require such
additional regularizations. In the context of coupled channels dynamics the
so-called chiral unitary approach has been widely used for meson-meson and
meson-baryon scattering \cite{coupl}. In a recent work \cite{oller}, Oller has
adopted the method to elastic nucleon-nucleon scattering, considering so far
only contact interactions. Clearly, a more realistic treatment must include
the one-pion exchange as well as pion-loop contributions. If one goes now in
this novel unitarization scheme for NN-scattering to sufficiently high orders 
in the small momentum expansion one will encounter the (multiply) iterated 
pion-exchanges. At present, analytical expressions are only known for the 
once-iterated pion-exchange scattering amplitude (see sect. 4.3 in
ref.\cite{nnpap}), but not for the higher iterations or the iterations
involving the pion-loop potential.  

The purpose of the present paper is to perform such calculations for the 
technically simpler case of a scalar-isovector boson (treating the nucleons in
the nonrelativistic approximation). The generalization to the pseudoscalar 
pion with its spin- and momentum-dependent couplings to the nucleon is in
principle straightforward and will be presented elsewhere \cite{murcia}
together with results for NN-phase shifts and mixing angles. For the following
we assume that the scalar-isovector boson of mass $m$ is light ($m\ll M$,
where $M$ denotes the nucleon mass) and that it couples weakly to the nucleon
(with a coupling constant $g\sim 2\pi m/M$).  
\begin{figure}
\begin{center}
\includegraphics[scale=1.0,clip]{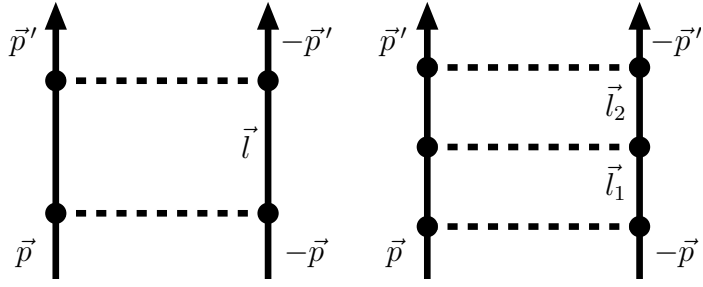}
\end{center}
\vskip -0.5cm
\caption{Diagrams of once-iterated (left) and twice-iterated (right)
boson-exchange}
\end{figure}
Let us start with recalling the once-iterated meson-exchange amplitude. The
corresponding planar one-loop diagram for the elastic scattering process in the
center-of-mass frame is shown in Fig.\,1. The pertinent complex-valued 
one-loop integral reads:
\begin{eqnarray} &&  \int {d^3 l\over (2\pi)^3}
{g^4 M (\vec \tau_1 \cdot \vec \tau_2)^2 \over (\vec l^{\,2} 
-\vec p^{\,2}-i0)[m^2 +(\vec l+\vec p\,)^2][m^2 +(\vec l+\vec p\,')^2]} \\ 
\nonumber &&= {g^4 M \over 4\pi} (3-2\vec \tau_1 \cdot \vec \tau_2)
\, G(p,q) \,, \end{eqnarray}
and it can be solved in terms of inverse trigonometric and logarithmic 
functions \cite{nnpap}: 
\begin{eqnarray} G(p,q) &=& {1\over q \sqrt{m^4 +p^2(4m^2+q^2)}} \Bigg[
\arcsin{  q \,m \over \sqrt{(m^2 +4p^2)(4m^2 +q^2)}} \\ \nonumber &&
+ \, i \, \ln{ p\, q + \sqrt{m^4 +p^2(4m^2+q^2)} \over m \sqrt{m^2 +4p^2}}
\Bigg] \,. \end{eqnarray} 
\begin{figure}
\begin{center}
\includegraphics[scale=0.53,clip]{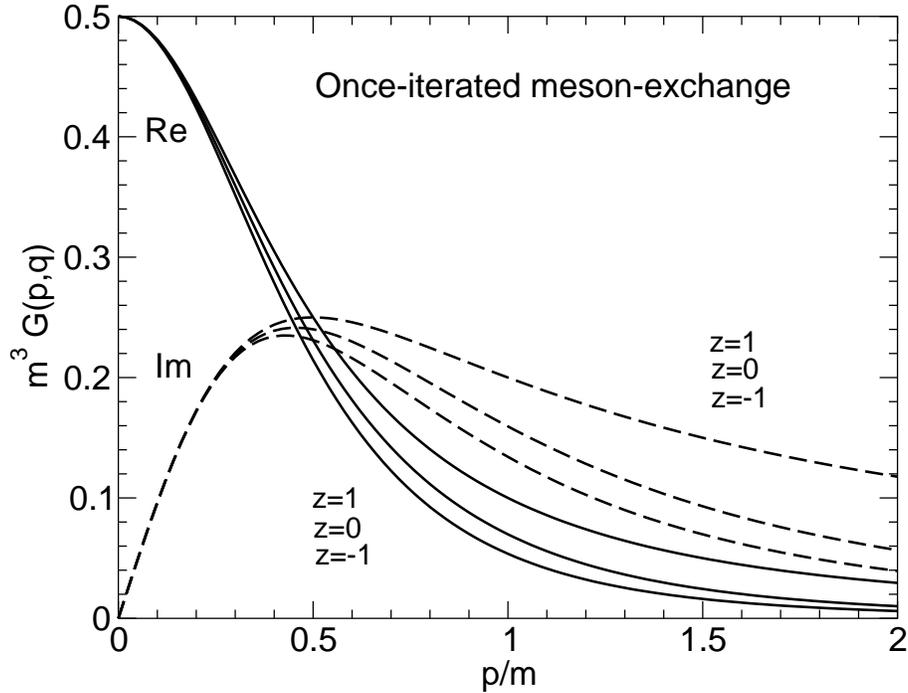}
\end{center}
\vskip -0.7cm
\caption{Real and imaginary part of the once-iterated scalar-boson 
exchange NN-scattering amplitude $G(p,q)$ versus the center-of-mass momentum
$p$. The cases $z=1,0,-1$ correspond to forward, perpendicular, and backward
scattering with momentum transfer $q= p \sqrt{2(1-z)}$, respectively.}
\end{figure}
Here, $p= |\vec p\,| =|\vec p\,'|$ is the center-of-mass momentum and $q = | 
\vec p\,' -\vec p\,|$ denotes the momentum transfer between the two nucleons.
$\vec \tau_{1,2}$ are the usual isospin operators. For the value at threshold 
$p=q=0$ one  easily deduces from Eq.(2): $G(0,0)= 1/(2m^3)$.  

In Fig.\,2 we show the real and imaginary part of the (dimensionless)
once-iterated scalar-boson exchange amplitude $m^3 G(p,q)$ as a function of
the center-of-mass momentum $p$ in the region $0\leq p \leq 2m$.  The three
full and three dashed lines correspond to the cases of forward scattering 
($z=1$), perpendicular scattering ($z=0$), and backward scattering ($z=-1$) 
with the momentum transfer given by $q= p \sqrt{2(1-z)}$. All 
other possible kinematical configurations lie of course in between. One 
observes from Fig.\,2 a rapid decrease of the real part, whereas the imaginary
part develops a broad  maximum around $ p \simeq 0.5 m$.      

Next, we come to the twice-iterated meson-exchange represented by the right
diagram in Fig.\,1. The corresponding two-loop integral:
\begin{eqnarray} && \int {d^3 l_1 d^3 l_2 \over (2\pi)^6} {g^6 M^2 (\vec 
\tau_1 \cdot \vec \tau_2)^3 \over (\vec l_1^{\,2}-\vec p^{\,2})(\vec l_2^{\,2}
-\vec p^{\,2})[m^2 +(\vec l_1+\vec p\,)^2] [m^2 +(\vec l_1-\vec l_2)^2][m^2 
+(\vec l_2+\vec p\,')^2]} \\ \nonumber &&= {g^6 M^2 \over 16\pi^2} (7\vec 
\tau_1 \cdot \vec \tau_2-6)  \, H(p,q) \,, \end{eqnarray} 
defines the complex-valued scattering amplitude $H(p,q)$. In order to evaluate
it, we make use of (perturbative) unitarity in the form of the Cutkosky
cutting rule. For the problem at hand it states that the imaginary part of the 
twice-iterated boson-exchange amplitude is equal to the two-body phase space 
integral (i.e. a solid angle integral) over the one-boson exchange amplitude 
times twice the real part of the once-iterated boson-exchange amplitude. Since
the latter is known in closed form (see Eq.(2)) we get after some
transformations the following single-integral representation for the imaginary
part:  
\begin{eqnarray} {\rm Im}\, H(p,q) &=& \int_0^{2p} dk \, \arcsin{ k\, m
\over \sqrt{(m^2 +4p^2)(4m^2 +k^2)}}\, \bigg\{ \Big[ m^4 +p^2(4m^2+k^2)\Big] 
\\ \nonumber && \times \Big[ m^4p^2 +m^2(2p^2(q^2+k^2) -q^2k^2)+p^2 (q^2-
k^2)^2\Big]\bigg\}^{-1/2} \,. \end{eqnarray}
The associated real part is obtained from an unsubtracted dispersion relation 
of the form:  
\begin{equation} {\rm Re}\, H(p,q)= {2 \over \pi} -\!\!\!\!\!\! \int_0^\infty
{dp'\,p' \over p'^{\,2} -p^2 }\, {\rm Im}\, H(p',q) \,. \end{equation}
It is astonishing that the value at threshold $p=q=0$ is still calculable 
analytically, with the result:
\begin{equation} m^4  H(0,0) = \ln{4 \over 3} \,. \end{equation} 
In case of the two-loop representation Eq.(3) we can show some intermediate 
steps: 
\begin{eqnarray} m^4 H(0,0) &=& {2\over \pi^2} \int_0^\infty dx
\int_0^\infty dy \int_{-1}^1 dz \Big[ (1+x^2)(1+y^2)(1+x^2+y^2-2xyz)\Big]^{-1}
\\ \nonumber &=&  {2\over \pi} \int_0^\infty {dx \over x(1+x^2)}\bigg[ \arctan
x- \arctan{x\over 2} \bigg] = \ln 2 - \ln{3\over 2}  \,, \end{eqnarray} 
which lead to this results. For the numerical evaluation of the principal 
value integral in Eq.(5) it is advantageous to  convert it into a sum of 
nonsingular integrals by the identity:
\begin{equation}  -\!\!\!\!\!\! \int_0^\infty ds'\,{f(s') \over s'-s }
= \int_0^{2s} ds'\,{f(s') -f(s)\over s'-s } +\int_{2s}^\infty ds'\,{f(s') 
\over s'-s } \,,  \end{equation}
where $s=p^2$. In Fig.\,3 we show the real and imaginary part of the 
(dimensionless) twice-iterated scalar-boson exchange amplitude $m^4 H(p,q)$
versus the center-of-mass momentum $p$ in the interval $0\leq p \leq 2m$. In 
comparison to the once-iterated boson-exchange amplitude $G(p,q)$ one observes
less dependence on the scattering angle, a feature which is exhibited by the
weaker splitting of the curves ($z = 1$) and ($z=-1$) corresponding to forward
and backward scattering. Another remarkable property is that the real part
Re\,$H(p,q)$ crosses zero at $p \simeq 0.58\,m$ and from there on it continues 
with small negative values which asymptotically tend to zero.     
\begin{figure}
\begin{center}
\includegraphics[scale=0.53,clip]{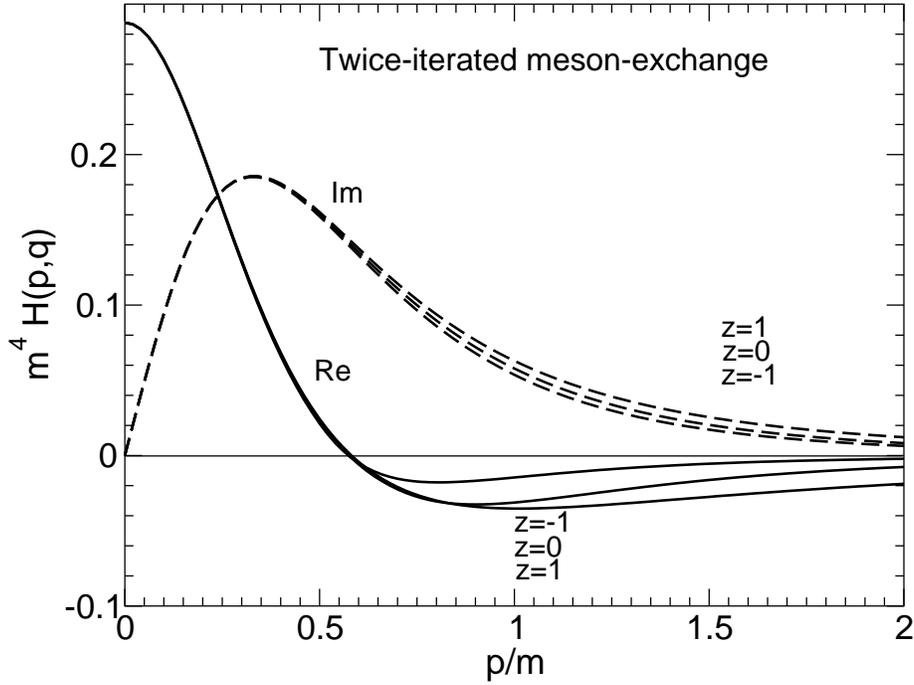}
\end{center}
\vskip -0.7cm
\caption{Real and imaginary part of the twice-iterated scalar-boson exchange 
NN-scattering amplitude $H(p,q)$. For further notations, see Fig.\,2.}
\end{figure}
\begin{figure}
\begin{center}
\includegraphics[scale=1.0,clip]{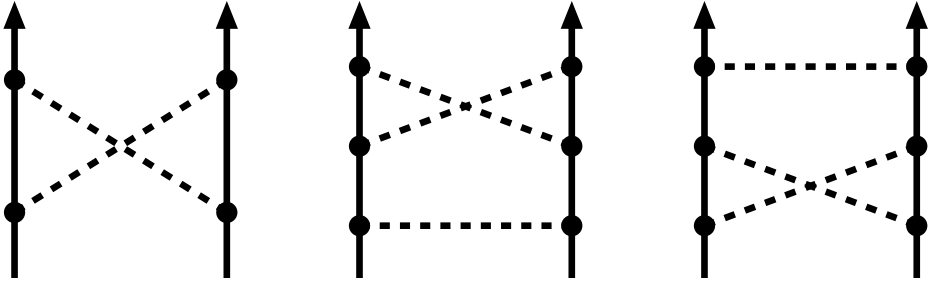}
\end{center}
\vskip -0.5cm
\caption{Crossed two-boson exchange diagram, and diagrams related
(partly) to the iteration of the irreducible one-loop potential $V_2(q)$ with
one-boson exchange.}
\end{figure}
At two-loop order there are also the non-planar ladder diagrams shown in 
Fig.\,4. These correspond (partly) to the iteration of the irreducible 
two-boson exchange (i.e. the one-loop potential) with the one-boson exchange. 
The complete one-loop potential arises from the (left) crossed two-boson 
exchange diagram in Fig.\,4 and the irreducible part of the (left) box diagram 
in Fig.\,1. Both pieces together lead to the following result for the (static) 
one-loop potential\,\footnote{In this work, we do not consider relativistic 
$1/M$-corrections to the irreducible one-, two- and three-boson exchange 
potentials.} in momentum space:     
\begin{equation} V_2(q) = {2g^4 \over \pi^2} \,\vec \tau_1 \cdot \vec \tau_2
\, {1 \over q \sqrt{4m^2+q^2}} \ln { q+ \sqrt{4m^2+q^2}\over 2m} \,.
\end{equation} 
Our sign convention is chosen here such that the (static) one-boson exchange 
reads: $V_1(q) = g^2 \, \vec \tau_1 \cdot \vec \tau_2\,(m^2+q^2)^{-1}$. It is 
interesting to note here that modulo their isospin factors $3 \pm 2\vec \tau_1
\cdot \vec \tau_2$ the crossed box diagram and the irreducible part of the
planar box diagram are equal but with opposite sign in the limit $M\to \infty$ 
(for further details on that fact, see sect.\,4.2 in ref.\cite{nnpap}). The 
coordinate space potential corresponding to Eq.(9) can be expressed through a
modified Bessel function: $\widetilde V_2(r) = -g^4 \vec \tau_1 \cdot \vec 
\tau_2\, K_0(2mr)/2  \pi^3 r$.

Now we are in the position to calculate also the iteration of the (static) 
one-loop two-boson exchange potential with the one-boson exchange: 
\begin{equation} T_3^{\times}(p,q) = {g^6 M \over 2\pi^3} (3-2 \vec \tau_1 
\cdot \vec \tau_2) \, J(p,q) \,. \end{equation}
Unitarity determines again the imaginary part of the complex-valued amplitude 
$J(p,q)$ as:
\begin{eqnarray} {\rm Im}\, J(p,q) &=& \int_0^{2p} dk\,\ln{k+\sqrt{4m^2+k^2}
\over 2m} \,\bigg\{(4m^2+k^2)\\ \nonumber && \times \Big[ m^4p^2 +m^2(2p^2(q^2 
+k^2)-q^2k^2)+p^2  (q^2-k^2)^2\Big]\bigg\}^{-1/2} \,, \end{eqnarray}
and the real part Re\,$J(p,q)$ follows from an unsubtracted dispersion 
relation analogous to Eq.(5). The value at threshold $p=q=0$ is again
calculable analytically, with the result: 
\begin{equation} m^3 J(0,0)=  \int_2^\infty {2\,dx \over x(x+1)\sqrt{x^2-4}} =
\pi \bigg( {1\over 2} - {2 \sqrt{3} \over 9} \bigg) \,, \end{equation} 
where we have made use of the spectral function representation \cite{nnpap} of 
the (static) one-loop potential $ V_2(q)$ written in Eq.(9).
\begin{figure}
\begin{center}
\includegraphics[scale=0.53,clip]{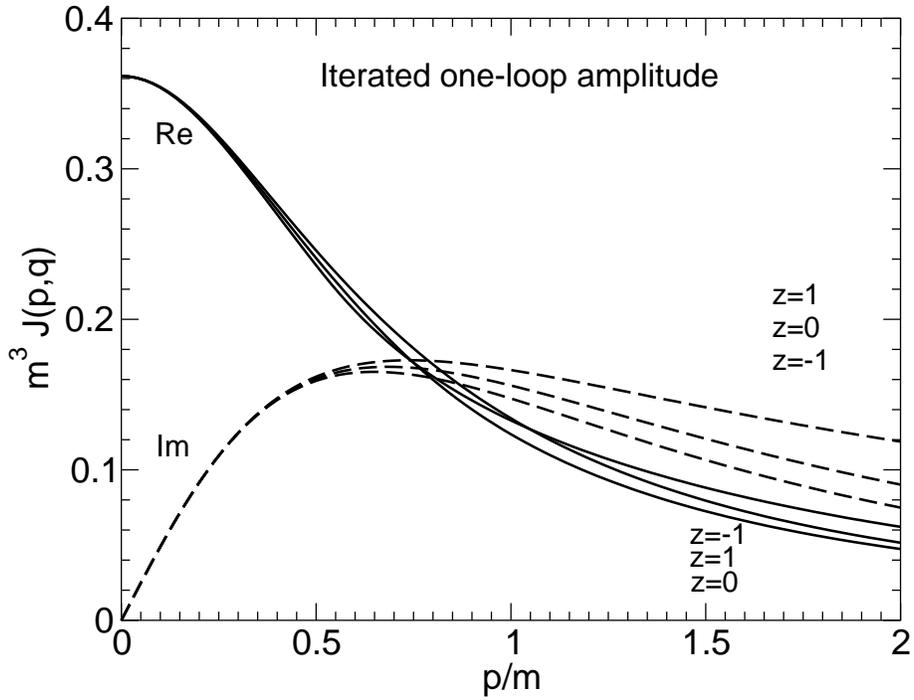}
\end{center}
\vskip -0.7cm
\caption{Real and imaginary part of the NN-scattering amplitude $J(p,q)$
generated by iterating the (static) irreducible one-loop potential with the 
one-boson exchange. For further notations, see Fig.\,2.}
\end{figure}
In Fig.\,5 we show the real and imaginary part of the (dimensionless) 
iterated one-loop amplitude $m^3 J(p,q)$ versus the center-of-mass momentum 
$p$ in the region $0\leq p \leq 2m$. The real part displays only a minor 
dependence on the scattering angle, which is more pronounced for the imaginary
part. Another noticeable feature is that the real part stays positive 
throughout and that the decrease of the amplitudes with increasing momentum 
$p$ is slower in comparison to $G(p,q)$ and $H(p,q)$ shown Figs.\,2,3.
The latter property has to do with the presence of the larger mass scale $2m$ 
in the (static) one-loop potential $V_2(q)$.     
\begin{figure}
\begin{center}
\includegraphics[scale=1.3,clip]{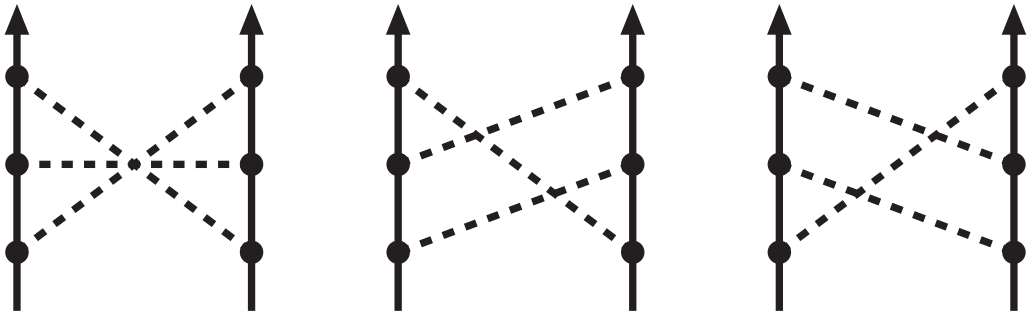}
\end{center}
\vskip -0.5cm
\caption{Crossed three-boson exchange diagrams. The dashed lines represent the
isovector-scalar boson of mass $m$.}
\end{figure}
At two-loop order there are in addition the three crossed three-boson exchange 
diagrams shown in Fig.\,6. These build up together with the irreducible parts 
of the two-loop ladder diagrams in Figs.\,1,4 the (irreducible) three-boson 
exchange potential. The separation of a two-loop ladder diagram into 
twice-iterated $(\sim M^2)$, once-iterated $(\sim M^1)$, and nonrelativistic 
irreducible components $(\sim M^0)$ is defined by the expansion of the
corresponding Feynman amplitude in powers of the reciprocal nucleon mass $1/M$.
For that one integrates first the product of light boson propagators $(l_j^2-
m^2+i0^+ )^{-1}$ and heavy nucleon propagators $2M(p_j^2 -M^2+i0^+)^{-1}$ over
the loop-energies $l^0_{1,2}$ via residue calculus and then expands in powers
of $1/M$. This way we find that modulo their isospin factors $7\vec \tau_1\cdot
\vec \tau_2 \pm 6$ the irreducible part of the right (3-rung ladder)
diagram\,\footnote{The systematic expansion in powers of $1/M$ gives rise also
to a (complex-valued) piece of order $M^0$ which is to be interpreted a
relativistic  $1/M^2$-correction to the twice-iterated boson-exchange.} in
Fig.\,1 agrees  with that of the left diagram in Fig.\,6 \cite{2looppot}. 
Discarding again the  isospin factors $7\vec \tau_1\cdot \vec
\tau_2 \pm 6$ and $-\vec \tau_1 \cdot \vec \tau_2\pm 6$ we deduce furthermore
that the total sum of the three crossed three-boson exchange diagrams in 
Fig.\,6 is equal to the difference between the right (middle) crossed diagram
in Fig.\,6 and the irreducible part of the right (middle) ladder diagram in
Fig.\,4. Putting back the isospin factors and using the four
relations between the  irreducible parts of the six two-loop diagrams, we find
for the isoscalar  component of the (static) three-boson exchange  potential
in  momentum space:     
\begin{equation} V_3(q)^{ (is)} = -g^6\int {d^3l_1 d^3 l_2 \over (2\pi)^6}\, 
{ 12  \over \omega_1^2 \omega_2^4 \omega_3^2} = - {3 g^6 \over 8 \pi^2 m
\,q } \, \arctan{q \over 3m} \,, \end{equation}
where we have made use of the results derived in the appendix of
ref.\cite{2looppot}. The quantities $ \omega_j = \sqrt{m^2 + \vec l_j^{\,2}}$
denote the on-shell energies of the three exchanged scalar-isovector bosons, 
where $\vec l_3 = \vec l_1 +\vec l_2+\vec q$. The corresponding potential in 
coordinate space is repulsive and can be expressed by a simple exponential
function: $\widetilde V_3(r)^{ (is)} =3g^6e^{-3m r}/(32\pi^3mr^2)$.  The 
isovector part of the (irreducible) three-boson exchange potential
proportional to $\vec \tau_1 \cdot \vec \tau_2$ reads on the other hand: 
\begin{equation} V_3(q)^{ (iv)} =  g^6\int {d^3l_1 d^3 l_2 \over (2\pi)^6}\, {4
\over \omega_1^3 \omega_2^3 \omega_3} \bigg[ {2\omega_2 \over (\omega_1+ 
\omega_3)^2} + {1\over \omega_1+\omega_2+\omega_3} \bigg]  \,. \end{equation}
 
At zero momentum transfer $q=0$ the six-dimensional integral in Eq.(14) can 
be evaluated analytically and the result $(6\pi m )^{-2}$ is found. For a
numerical evaluation of the isovector potential $V_3(q)^{ (iv)}$ the 
representation in Eq.(14) is not best suited. A better representation is
obtained by introducing Feynman parameters for each four-dimensional loop 
integral and performing the elementary integrations over some of the Feynman
parameters. After several skilful transformations we arrive at the following 
very handy double-integral representation:       
\begin{eqnarray}  V_3(q)^{ (iv)} &=& {g^6 \over 4\pi^4}\int_0^1\!\!dx 
\int_0^\infty \!\! dy \,\bigg\{1-\sqrt{y}\arctan{1\over \sqrt{y}}\bigg\}
\\ \nonumber &&\times \Big[m^2(x^2+x y+ y+y^2)+ q^2 x(1-x)y\Big]^{-1}  \,. 
\end{eqnarray} 
In this form the $dx$-integral could even be solved in
terms of square-root and arctangent or logarithmic functions (depending on the
sign of a radicand). The associated coordinate space potential $\widetilde
V_3(r)^{(iv)}$ can now also be computed easily. One just has to replace the 
polynomial inside the square brackets in Eq.(15) by the expression $-4\pi r x 
(1-x)y \exp\{ m r[(x^2+x y+ y+y^2)/(x(1-x)y)]^{1/2}\}$. In the first and
second row of Table\,I we have collected numerical values which display the 
dropping of the isoscalar and isovector three-boson exchange potentials
$V_3(q)^{(is,iv)}$ with the momentum transfer $q$. In each case we have
divided $V_3(q)^{(is,iv)}$ by the corresponding value at $q=0$. For
comparison, we give in the third row the analogous ratios for the one-loop
two-boson exchange potential $V_2(q)$ written down in Eq.(9). One observes
that the isovector three-boson exchange potential drops somewhat faster than
its isoscalar counterpart. This decrease is however weak in comparison to that
of the two-boson exchange potential, mainly due to the different mass scales
($3m$ versus $2m$) involved. As an aside, we note that for a scalar-isoscalar
boson the  (static) irreducible two-boson and three-boson exchange potentials 
would both vanish identically: $V_2(q)= V_3(q)=0$. The latter zero-result 
follows from the relations between the irreducible parts of the six
three-boson exchange two-loop diagrams, mentioned above. 

\bigskip
\begin{table}[hbt]
\begin{center}
\begin{tabular}{|c|cccccccccc|}
    \hline
$q/m$ & 1 & 2 & 3 & 4 & 5 & 6 & 7 & 8 & 9 & 10 \\
   \hline
 3-isoscalar & 0.965 & 0.882 & 0.785 & 0.695 & 0.618 & 0.554
   & 0.500 & 0.455 &0.416 & 0.384 \\
   3-isovector & 0.957 & 0.856 & 0.739 & 0.633 & 0.544 & 0.471
   & 0.412 & 0.363 &0.322 & 0.289 \\ one-loop & 0.861 & 0.623 & 0.442 & 0.323 
& 0.245 & 0.192   & 0.154 & 0.127 &0.107 & 0.091 \\ 
2-isoscalar & 0.927 & 0.785 & 0.655 & 0.554 & 0.476 & 0.416
   & 0.369 & 0.331 &0.300 & 0.275 \\
   2-isovector & 0.960 & 0.874 & 0.783 & 0.704 & 0.637 & 0.581
   & 0.534 & 0.494 &0.460 & 0.430 \\
\hline
\end{tabular}
\end{center}
{\it Table I: First and second row: The isoscalar and isovector three-boson 
exchange potentials Eqs.(13,15) versus the momentum transfer $q$. The 
respective values at $q=0$ have been divided out. The third row gives the 
analogous ratios for the one-loop two-boson exchange potential Eq.(9). The 
fourth and fifth row correspond to the two-boson-exchange potentials with
one-loop vertex corrections written in Eqs.(18,19).}
\end{table}

\begin{figure}
\begin{center}
\includegraphics[scale=1.0,clip]{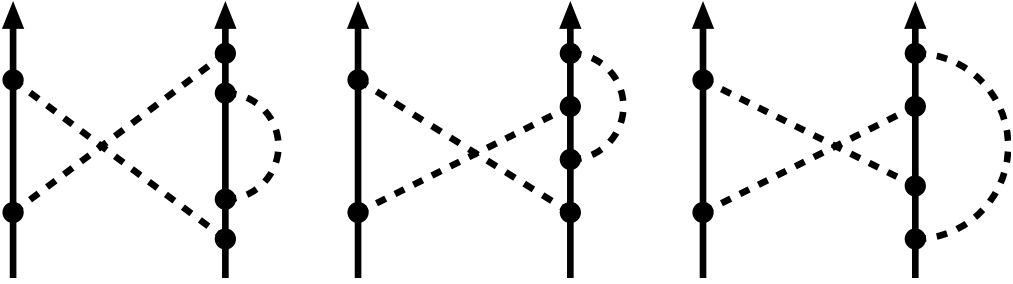}
\end{center}
\vskip -0.5cm
\caption{Two-boson exchange diagrams with one-loop vertex corrections. The
diagrams with uncrossed boson lines and the role of both nucleons interchanged
are not shown.}
\end{figure}

For the sake of completeness (at two-loop order in the nonrelativistic 
approximation) we consider also the two-boson exchange diagrams with vertex 
corrections shown in Fig.\,7. In order to evaluate them we make use of the 
methods outlined in ref.\cite{2loop}. From the one-loop corrections (beyond 
mass and coupling constant renormalization) to the isospin-even and
isospin-odd boson-nucleon scattering amplitudes:\footnote{For a
 scalar-isoscalar boson these one-loop corrections would vanish identically.} 
\begin{equation} T^+_{bN}(\omega) = {2g^4 \over \pi \omega^2} \Big( m-
\sqrt{m^2-\omega^2} \Big) \,, \qquad  T^-_{bN}(\omega) = {2g^4 \over \pi^2 
\omega^2} \bigg[ \omega -\sqrt{m^2-\omega^2}\arcsin{\omega \over m} \bigg] \,,
\end{equation} 
one can calculate (via unitarity) the spectral functions Im\,$V_2(i\mu)$. 
These mass spectra determine then the two-boson exchange potentials in momentum
space through an unsubtracted dispersion relation:   
\begin{equation} V_2(q) = {2\over \pi} \int_{2m}^\infty d\mu \, {\mu\, {\rm
Im} V_2(i\mu) \over \mu^2+q^2} \,. \end{equation}
We find for the isoscalar part of the  two-boson exchange potential with
one-loop vertex corrections: 
\begin{equation} V_2(q)^{(is)}_{vc} = {3g^6 \over 4\pi^2 m \, q} \arctan{q 
\over 2m} \,, \end{equation} 
whose associated coordinate space representation is a simple exponential 
function: $\widetilde V_2(r)^{(is)}_{vc}= -3g^6\, e^{-2m r}/(16 \pi^3 m r^2)$. 
The isovector component proportional to $\vec \tau_1 \cdot \vec \tau_2$ reads 
on the other hand: 
\begin{equation} V_2(q)^{(iv)}_{vc} = {2g^6 \over \pi^4 m \, q} \int_1^\infty
{dx\,x \over (x^2-1)^2 }\Big[ \sqrt{x^2-1} -x \ln(x+ \sqrt{x^2-1}) \Big]
\arctan{ q \over 2m x} \,, \end{equation}
with the value $-g^6/(2\pi^4 m^2)$ at zero momentum transfer $q=0$. In order
to obtain the associated coordinate space potential one just has to replace 
$q^{-1} \arctan(q/2mx)$ in Eq.(19) by the expression $-e^{-2mr x}/(4\pi r^2)$.
The numbers in the fourth and fifth row of Table\,I display the dependence
of these two-boson exchange potentials with one-loop vertex corrections 
$V_2(q)^{(is,iv)}_{vc}$ on the momentum transfer $q$. One observes a slow 
decrease similar to that of the irreducible three-boson exchange potentials.

In summary we have calculated in this work all nonrelativistic contributions 
from scalar-isovector boson-exchange between nucleons at one- and two-loop 
order (including vertex corrections). At one loop order these contributions 
consist of the once-iterated boson-exchange amplitude and the irreducible
two-boson exchange potential. At two-loop order one has the complex-valued
scattering amplitudes related to the twice-iterated boson-exchange and the
iteration of the one-loop potential with the one-boson exchange as well as the
irreducible three-boson exchange potentials and the two-boson exchange
potentials with one-loop vertex corrections. In each case we have given
analytical expressions involving only the minimum number of integrations (to
be done numerically). The present results could be useful for phase-shift
analyses, few- and many-body calculations, etc. The applied methods can be
straightforwardly generalized to the pseudoscalar-isovector pion with its
spin- and momentum-dependent couplings to the nucleon. The number of
amplitudes and diagrammatic contributions to each amplitude increases however
drastically by several orders of magnitude. Work along these lines is in
progress \cite{murcia}. In passing we have to point out that there exist still 
other two-loop contributions which also scale as $g^6 M^0$. These
complex-valued amplitudes correspond either to the iteration of the 
relativistic  $1/M$-correction to the one-loop potential with the one-boson 
exchange, or to relativistic $1/M^2$-correction to the twice-iterated 
boson-exchange. More work is necessary in order to isolate, classify, and
analytically compute these relativistic correction terms.

\end{document}